\documentclass[article]{aa}
\usepackage{graphics}
\usepackage{graphicx}
\usepackage{txfonts}
\begin{document} 
          \title{The $XMM-Newton$ $\Omega$ project: III. Gas mass fraction shape in high redshift clusters}  
 
   \titlerunning{Gas mass fraction in distant clusters}

   \author{
 R.~Sadat
\inst{1},
 A.~Blanchard
\inst{1},
S.C.~Vauclair
\inst{1},
D.H.~Lumb 
\inst{2},
J. Bartlett 
\inst{3},
 A.K. Romer
\inst{4},
J.-P.~Bernard
\inst{5},
M.~Boer
\inst{6},
 P.~Marty
\inst{7},
J.~Nevalainen
\inst{8},
D.J. Burke
\inst{8},
 C.A. Collins 
\inst{9},
R.C. Nichol
\inst{10}   
}

 \institute{
Laboratoire d'Astrophysique de Tarbes et Toulouse, OMP, CNRS, UMR 5572, UPS, 14 Avenue. E. ~Belin, F--31400 Toulouse, France
\and
Advanced Concepts and Science Payloads Office, European Space Agency, ESTEC, 2200AG Noordwijk, The Netherlands 
\and
Laboratoire de Physique Corpusculaire et Cosmologie, Coll\`ege de France, 11 pl. Marcelin Berthelot, F-75231 Paris Cedex 5, France 
\and
Astronomy Center, Department of Physics and Astronomy, University of Sussex Falmer, Brighton BN19QH, United Kingdom 
\and
Centre d'\'etude spatiale des rayonnements, OMP, UPS, 9, Av. du Colonel Roche, BP4346, 31028 Toulouse, France
\and
Observatoire de Haute Provence, F - 04870 Saint Michel l'Observatoire, France
\and
Institut d'Astrophysique Spatiale, b\^at 121, Universit\'e Paris Sud XI,
91~405 Orsay cedex, France
\and
Harvard-Smithsonian Center for Astrophysics, 60 Garden street, Cambridge, MA 02138, USA
\and
Astrophysics Research Institute, Liverpool John Moores University, Twelve Quays House, Egerton Wharf, Birkenhead CH41 1LD, UK 
\and
Physics Department, Carnegie Mellon University, Pittsburgh, PA 15213, USA
}

\offprints{R.Sadat}
\mail{rsadat@ast.obs-mip.fr}
   
\date{Received \rule{2.0cm}{0.01cm} / accepted \rule{2.0cm}{0.01cm} }
\authorrunning{Sadat \& al. }

\abstract{
We study  the gas mass fraction behavior  
in distant galaxy clusters observed within the $XMM-Newton$  $\Omega$ project.
The typical  gas mass fraction $f_{\rm gas}$ shape of high redshift galaxy 
clusters 
follows the global shape inferred at low redshift quite well, once scaled 
appropriately : the  gas mass fraction increases with radius and flattens 
outward. This result is consistent with the simple picture in which clusters 
essentially form by gravitational collapse, leading to self similar 
structures for both the dark and baryonic matter. However, 
 we find that the mean gas profile 
in distant clusters shows some differences to local ones, indicating a 
departure from strict scaling. Assuming an Einstein-de Sitter cosmology, 
we find a slight 
deficit of gas in the central part of high-$z$ clusters. 
This result is consistent with the observed evolution in the 
luminosity-temperature relation.
We quantitatively investigate this departure from scaling laws by comparing 
 $f_{\rm gas}$ from a sample of nearby galaxy clusters 
(Vikhlinin, Forman \& Jones, 1999) to our 
eight high-z clusters. 
Within the local sample, a moderate but clear variation of the
 amplitude of the gas mass fraction with
temperature is found, a trend that weakens in the outer regions. 
 Taking into account these variations with radius and temperature, the apparent scaled 
gas mass fractions
in our distant clusters still systematically differ from local clusters.
{ This reveals that the gas fraction  does not strictly  follow a scaling 
law with redshift.
This provides clues to understand  the  redshift evolution of 
the $L-T$ relation   
whose origin is probably due to non-gravitational 
processes during  cluster formation. An important implication 
of our results is that the gas fraction evolution, a test of the cosmological 
parameters, 
can lead to biased values when applied at radii smaller 
than the virial radius.} From our $XMM$ clusters,  as well as $Chandra$ 
clusters in the same redshift range,  the apparent gas 
fraction at the virial radius obtained by extrapolation of the 
inner gas profile
 is consistent with a non-evolving universal value 
 in a high matter density model while in a concordance,
 model high redshift clusters show  an apparent 
higher $f_{\rm gas}$ at the virial radius
than to local clusters. 
}


\maketitle

\section{Introduction}

     Clusters of galaxies are unique cosmological probes whose
 statistical properties represent major sources of information for understanding the
history of structure formation as well as for the determination of the cosmological 
parameters.  X-ray observations are particularly relevant in this perspective
as they allow one to estimate the distribution of both the baryonic and total mass components,
a rather unique situation when studying structures in cosmology. 
In the simplest picture of purely gravitationally-driven formation of virialized systems like galaxy clusters, it is expected that such objects exhibit self-similarity (Kaiser 1986). In this model physical properties of galaxy clusters obey scaling laws which naturally emerge from the fact that there is no 
preferred scale and therefore two clusters of different masses should 
have identical internal structure when normalized to the virial radius. 
Furthermore, such internal structure should be independent of redshift. Self-similarity applies to both the dark matter
component and to the hot X--ray emitting intra-cluster medium (ICM).
As clusters of different masses arise from fluctuations of different amplitude 
(relative to the r.m.s. value), such a scaling is not expected to hold exactly.
Furthermore, in cosmological models different from the Einsten-de Sitter model,
the strict self-similarity of the expansion of the universe might be broken. 
 Nevertheless numerical simulations have shown that the relations between 
physical quantities expected from the scaling laws  
hold very well (Bryan \& Norman 1998). Comparison of expected relations
to observations is therefore expected to provide key information on 
their formation processes. 

 The observed properties of clusters are different from the 
scaling predictions, for example the observations lead to a luminosity-temperature relation which scales as  $L \propto T^{3}$ while theoretical models predict $L \propto T^{2}$. Such 
deviations from scaling laws are interpreted as due to non-gravitational 
processes such as preheating  by early galactic winds 
(e.g. Kaiser 1991; Evrard \& Henry 1991, 
David et al. 1995, Cavaliere, Menci, \& Tozzi 1998) or to radiative cooling 
(Pearce
et~al. 2000; Muawong et~al. 2002) and feedback from star formation or AGN (Voit 
\& Bryan 2001, Valageas \& Silk 1999). The excess of entropy (the so-called 
''entropy floor'') in cold system, provides  further evidence of the 
importance of non-gravitational processes (Ponman et~al. 1999;
Lloyd-Davis et~al. 2000). Although some numerical simulations including 
radiative cooling and/or pre-heating were able to reproduce the observed 
steepening in the {\hbox{$L_{\rm X}$--$T$} } relation  consistent with the
observations (Bialek et~al. 2001, Borgani et~al. 2001), it is yet unclear 
whether the relevant physics  has been properly identified and implemented.\\

Self-similar models also make predictions on the evolution of cluster 
properties. In particular the {\hbox{$L_{\rm X}$--$T$} } relation should scale 
as $(z+1)^{\Gamma}$ where $\Gamma$ should be equal to 3/2 in an Einstein de 
Sitter (EdS) universe. Several studies have found evidence of a 
weak evolution in the {\hbox{$L_{\rm X}$--$T$} } relation (Sadat et~al. 1998;
  Reichart et~al.
 1999). However, 
the luminosity estimates depend on the assumed cosmological
parameters as does the constraint on the amount of evolution. From the analysis of recent $XMM-Newton$ data of high-z clusters  it has 
been found that $\Gamma \sim 0.65$ in an EdS Universe
 while in a concordance model this value is close to 1.5, { close to the} value expected 
accordingly to standard scaling laws (Lumb et al. 2004).
 This result is consistent with previous investigations based on ASCA and 
$Chandra$ data (Sadat et al. 1998; Novicki et al. 2002; Vikhlinin et al. 2002).
The cosmological 
implication of such evolution has been presented in Vauclair et al. (2003).
The aim of the present study is to better understand the evolution  
of the gas mass fraction with redshift. We will show that understanding these 
properties is important to put constraints on the
cosmological parameters by requiring that the gas mass fraction remains 
constant with look-back time. Indeed, comparing the profiles of clusters at
different redshifts provides more
information than simply considering global quantities such as the
total X-ray luminosity.

We base our analysis on the XMM data obtained on a sample of eight
distant clusters observed as part of the $XMM-Newton$ $\Omega$
project, a systematic $XMM-Newton$ guaranteed time follow-up of the
most distant SHARC clusters (Bartlett et al. 2002). The high
sensitivity of $XMM-Newton$ allows us to investigate emissivity in high
redshift clusters beyond half the virial radius, a remarkable result
(Arnaud et~al. 2002). Our sample represents an homogeneous sample of
eight bona fide clusters with median luminosities between 2 and 15
$10^{44}$ erg/s (in an Einstein de Sitter cosmology with a Hubble
constant of 50 km/s/Mpc) with redshifts in a relatively restricted
range, between 0.45 and 0.65. This sample is therefore expected to be
fairly representative of the cluster population at high redshift,
allowing a systematic analysis of the gas mass profiles and therefore
allowing us to address the issue of gas mass fraction self-similarity and
its implications in constraining the cosmological parameters. Moreover,
the high sensitivity of $XMM-Newton$ makes possible a  statistical 
investigation of the outer gas distribution in this sample, a
key aspect as we will see.  The detailed data reduction and analysis
of this sample is presented in Lumb et al. (2004).  The present paper
is organized as follows. In section 2 we discuss the expectation of
the gas mass fraction from scaling arguments as well as the results of
the comparison of gas mass fraction in distant clusters to low
redshift ones.  In section 3, we discuss the consequence of our
findings for the use of clusters as cosmological probes. Finally, our
conclusions are given in section 4.  We used a Hubble
constant of 50 km/s/Mpc unless the dependence is explicitly given (with
$H_0 = 100h$km/s/Mpc).

\section{Scaling properties}

\begin{figure*}[!ht]
\includegraphics[scale=0.4,angle=0]{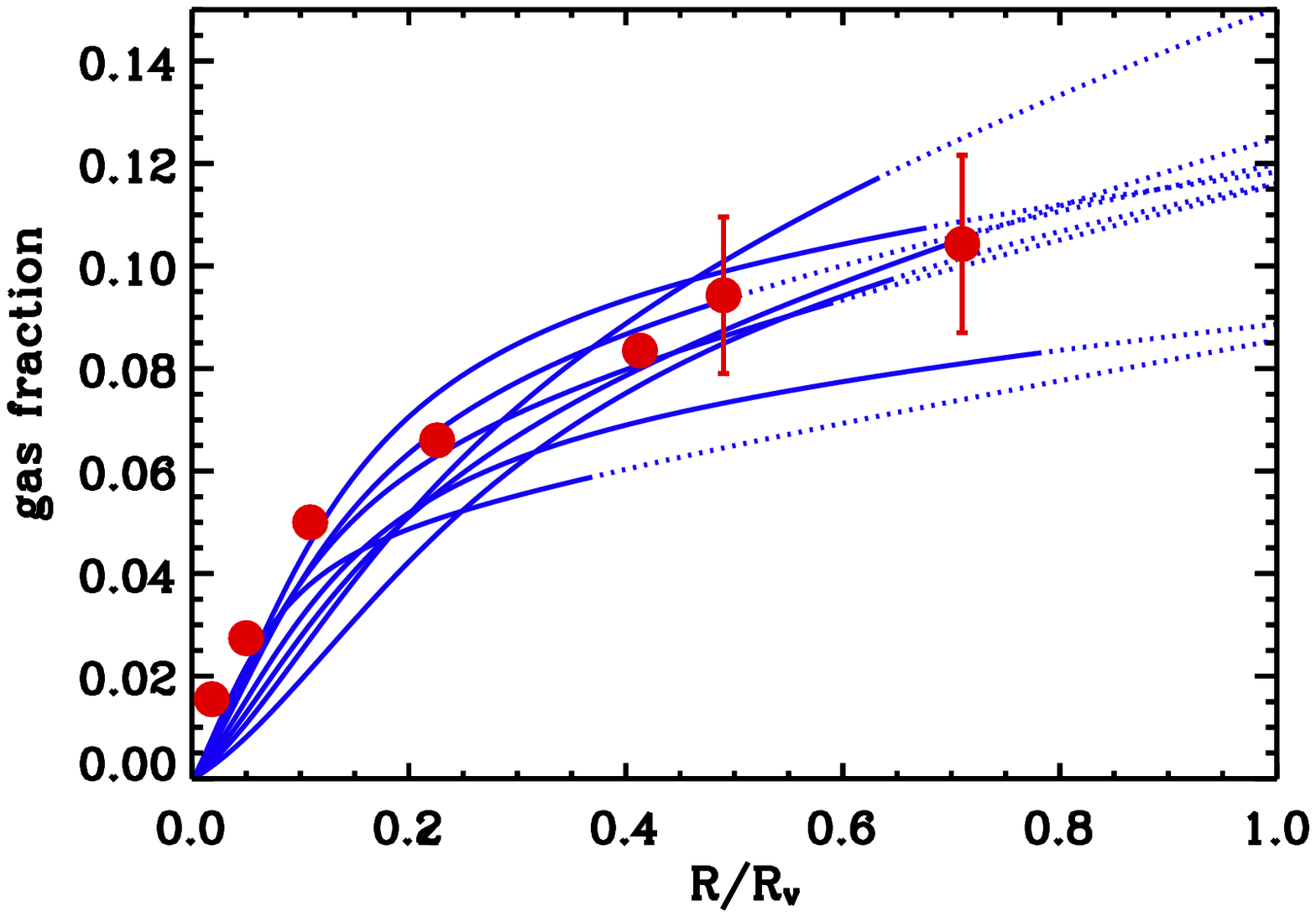}
\includegraphics[scale=0.4,angle=0]{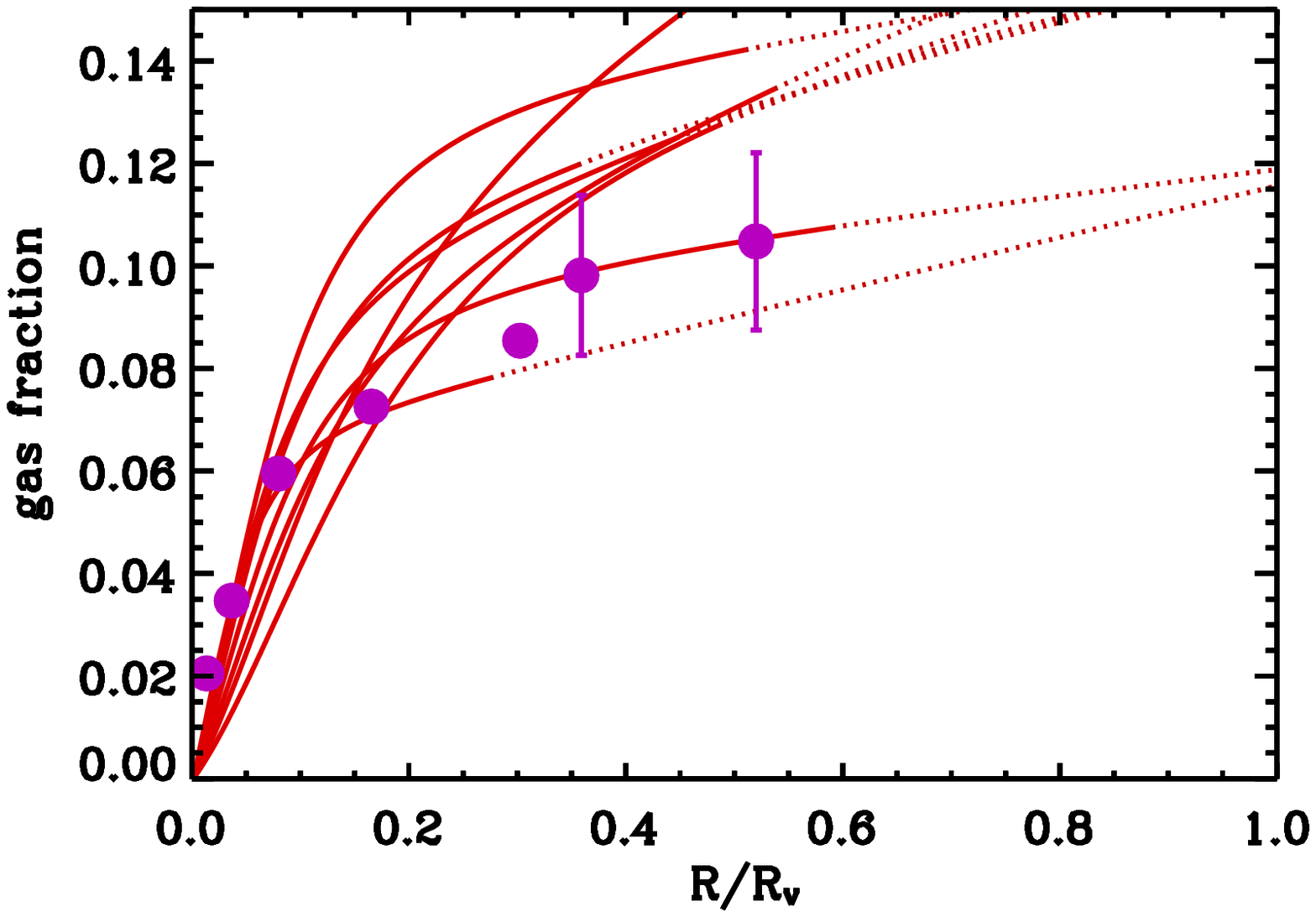}
\includegraphics[scale=0.4,angle=0]{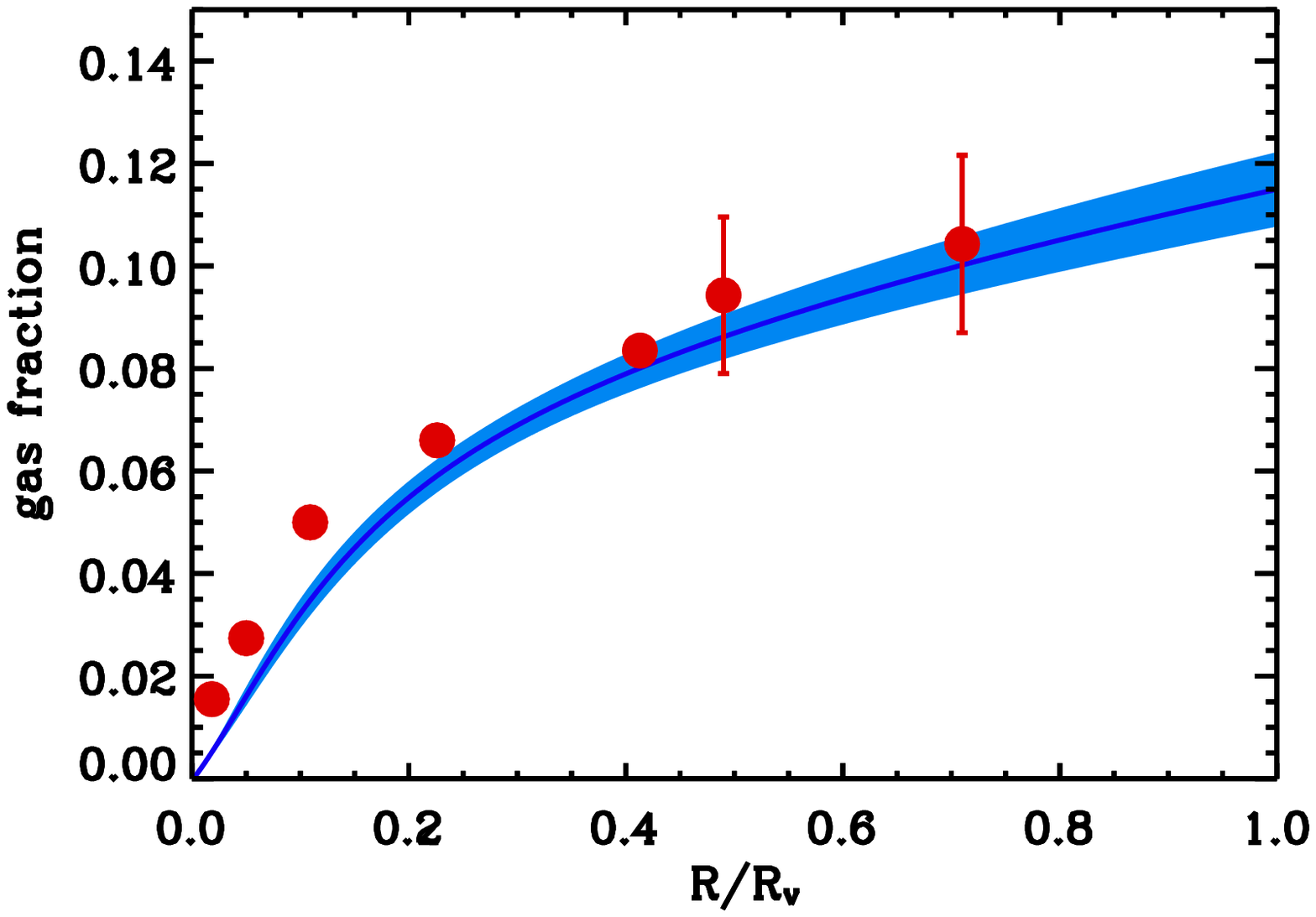}
\includegraphics[scale=0.4,angle=0]{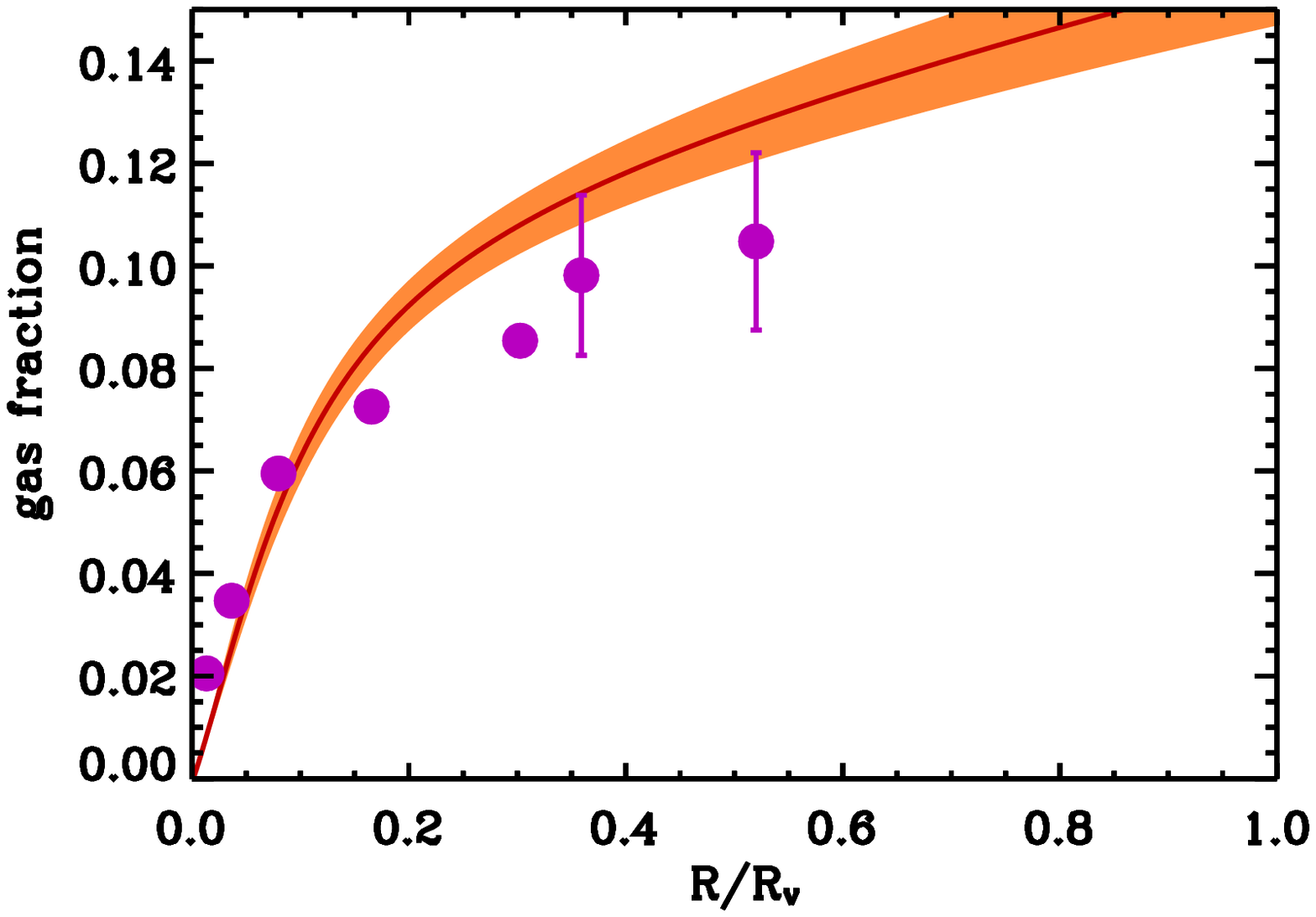}
\caption[]{Apparent $f_{\rm gas}$ plotted versus radius normalized to the virial
 radius 
$R/R_{V}$. The  left side corresponds to $f_{\rm gas}$ in an Einstein-de 
Sitter model and the right side  to the concordance model. In the upper 
graphs, the lines show the individual clusters $f_{\rm gas}$ up to  the maximum radius of detection (see Lumb et al. 2004), dots correspond 
to extrapolated $f_{\rm gas}$. The data (red
and purple circles) are  the $f_{\rm gas}$ in local sample (from RSB00 for the inner part and from VFJ99 in the outer parts) in an Einstein de Sitter cosmology  and in a flat low density model. The error bars 
correspond to  the typical dispersion in the VFJ99  sample. In the lower 
graphs, the average apparent $f_{\rm gas}$ and its uncertainty are plotted (filled area).}
\label{fig:fbshape}
\end{figure*}


Self-similar assumptions imply that the radial profile of any physical 
quantity should exhibit a similar shape independently of the cluster mass and at 
any 
redshift, once normalized to the virial radius. Numerical simulations 
in which only gravitational physics is taken into account indicate 
that halos of different masses follow
 a universal density profile, the so-called NFW profile 
(Navarro, Frenk \& White
 1996). On the observational side, the X-ray emissivity profiles 
in hot galaxy clusters have been also found to 
be very similar, at least in the outer part (Ponman et al. 1999; 
Neumann \& Arnaud 1999,  Arnaud, Neumann \& Aghanim 2001). Furthermore, evidence has been found that radial profiles of the $f_{\rm gas}$ , as well as the baryon fraction 
$f_b$, are  similar and seem
 to follow  a roughly  universal shape (Roussel et al. 2000, hereafter RSB00).
It has been found that such a universal profile is in reasonable agreement 
with the predictions of numerical simulations  (Sadat \& Blanchard 
2001; hereafter SB01). 
This supports the idea that the gas structure has not been strongly 
disturbed by non-gravitational processes and  supports  the principle 
of using their properties to constrain cosmological parameters.\\
Previous studies of the baryonic content in clusters indicated that baryons 
contribute  15-20\% of the total cluster mass (for $h = 0.5$); if the baryon 
fraction $ f_{\rm b} = M_{b}/M$ is representative of the universe as a whole
and, provided that  the actual baryon abundance is known,   
the cosmological matter density parameter $\Omega_M$ should lie in the range 
 $ \Omega_{\rm b}/f_{\rm b} =0.2-0.5$ 
(White et al. 1993, David et al. 1995, Evrard 1997).
However, although the gas mass fraction profile follows quite well the 
self-similarity assumption (RSB00), the density parameter 
derived from the baryon fraction 
estimation  might be corrupted by different effects that are related 
to the internal structure of the gas and that 
could alter the inferred value  (SB01). 

\subsection{Estimation of the cluster $f_{\rm gas}$ under the scaling hypothesis}

As already mentioned, the self-similarity hypothesis implies that the spatial variation of 
any physical quantity depends solely on $R/R_{\rm v}$; the virial  radius $R_{\rm v}$
can be  obtained from  its definition $M_{\rm v} = 4/3\pi \overline{\rho}(1+z)^3(1+\Delta_{\rm v}) R_{\rm v}^3$:
\begin{equation}
R_{\rm v} =  1.34 M^{1/3}_{15}(\Omega_0 (1+\Delta_{\rm v})/179)^{1/3}h^{-2/3}(1+z)^{-1} (h^{-1}\rm Mpc)
\label{eq:rv}
\end{equation}
 where $\Delta_{\rm v}$ is the virial contrast density compared to the universe. In an Einstein-de Sitter 
universe, one has $\Delta_{\rm v} = 18\pi^2$. In other cosmological models, there is 
some ambiguity as to how to define the proper reference radius for scaling 
relations. Commonly, the virial radius is defined from the spherical top-hat
model (see for instance Bryan \& Norman, 1998; note that these authors 
provide useful fits to compute $\Delta_{\rm v}$ with formulae involving $\Delta_c$, 
the contrast density, compared to a critical universe 
of density $\rho_c(z) = 3H(z)^2/8\pi G$). The
scaling of the mass-temperature relation is then obtained from $T \propto
GM/R_{\rm v}$: 
\begin{equation}
T =A_{TM} M^{2/3}_{15}(\Omega_0 (1+\Delta_{\rm v})/179)^{1/3}h^{2/3}(1+z)\rm keV
\label{eq:tm} 
\end{equation}
The normalization $A_{TM}$ can be obtained from  
numerical simulations (Evrard, Metzler, Navarro 1996, Bryan \& Norman 1998)
 or inferred from observations.
In the following, we use the same calibration as in 
SB01,  $A_{TM}= 5.86$, allowing a direct comparison with their results. Several studies have been performed in order to test the {\hbox{$M_{\rm V}$--$T$} } 
relation as predicted from numerical simulations by means of X-ray observations
 (see e.g. Horner et~al. 1999, Nevalainen et~al. 2000, Finoguenov et~al. 2001,
 Sanderson et~al. 2003). Disagreements have been found concerning both the 
normalization $A_{TM}$ and the slope (steeper than the predicted 1.5) for 
cooler systems (${T_{\rm X}}$ less than 4  keV). Note however that different 
normalizations  $A_{TM}$ of the {\hbox{$M_{\rm V}$--$T$} } relation are not 
expected to make a difference in the comparison
between local and high redshift samples, therefore the $f_{\rm gas}$ test 
is  essentially based on the  assumption that scaling of the $M_{\rm gas}-T$ relation
 is correct.\\ 
The gas mass fraction at a given radius $f_{gas}(r)=M_{gas}(r)/M_{tot}(r)$ is 
computed for each cluster. The gas mass profile follows directly from the 
electron number density profile: 
\begin{equation}M_g(r) = 4 \pi  \frac{m_p}{1-Y/2.}\,\int\limits_0^{r} n_e(r)\,  
r^2\, dr.\label{eq:mgas}
\end{equation}
 where $Y$ is the helium mass fraction (hereafter $Y = 0.25$). We assume a fully ionized gas,  spherically distributed, and a 
$\beta$--model for its distribution, which is known to provide a good 
representation of the gas out to the outer regions (VFJ99):
\begin{equation}
   n_{\rm e}(r) =   n_e(0) \left( 1 + \left( \frac{r}{{r_{\rm cX}}} \right)^2
   \right)^{-\frac{3}{2} \beta}
\end{equation}
where ${r_{\rm cX}}$ end $\beta$ corresponding to the 
best fit values derived in Lumb et al. (2004). The central gas density $n_e(0)$ was derived from the normalization K of the XSPEC Mekal model defined by

\begin{equation}
{K} = \frac{10^{-14}}{4 \pi (D_a*(1+z))^2} \int n_e n_p dV.
\label{eqn:K}
\end{equation}
where  $n_e$ and $n_p$ are respectively the electron and proton number 
densities and $D_a$ is the  angular  distance to the cluster. We assume 
$n_{\rm p} = 0.82 n_{\rm e}$ in the ionized intra-cluster plasma. The emission 
integral   $ EI \propto \int n_e n_p dV $ was 
evaluated   assuming that the x-ray emission extends up to the virial radius, 
in order to be fully consistent with Lumb et al. (2004). 
There is some arbitrariness in the assumption of the radius up to which the emission has 
to be taken into account, ranging from the largest radius at which the
 emission is detected up to infinity. For our sample, the contribution
to the flux of the emission, estimated by extrapolation of the fitted profile,
 beyond the detection radius is less than 1\%, 
and therefore represents a negligible source of uncertainty on our derived 
gas masses.

The normalization K value of each cluster was  taken from Table 2 of 
Lumb et al. (2004). \\
The dark matter profile was assumed to follow the NFW analytical profile 
(Navarro et~al. 1996) with a concentration parameter $c = 5$  in  order to 
allow direct comparison with  RSB00 (again, changing the value of the 
concentration parameter
is not expected to modify  the relative comparison of local and distant 
clusters).
In this study we will consider two cosmologies: an Einstein-de-Sitter (EdS) 
Universe ($\Omega_m =1$) and a concordance model ($\Lambda$CDM) Universe with
($\Omega_m =0.3,\Omega_\lambda=0.7$). The apparent $f_{\rm gas}$ values 
of our $XMM-Newton$ clusters estimated at ${r_{\rm 500}}$ are given in Table 
\ref{tab:fg} for both cosmologies. The mean $f_{\rm gas}$ value at 
${r_{\rm 500}}$ is $f_{\rm gas} = 0.095$ in an EdS  model and 
$f_{\rm gas} = 0.14$ in a $\Lambda$CDM model.   
\begin{table*} 
\begin{tabular}{l l l l l l l l l}\hline\hline 
 Cluster  name & RXJ0337.7&RXJ0505.3&RXJ0847.2&RXJ1120.1&RXJ1325.5&RXJ1334.3&RXJ1354.2&RXJ1701.3\\ \\\hline 
$z$ &0.577 &0.51&0.56&0.60&0.445&0.62&0.551&0.45\\ \\\hline  
T (keV)& 2.6 & 2.5 & 3.62 & 5.45 & 4.15 & 5.2 & 3.66 & 4.5\\  
& $^{+0.4}_{-0.3}$ & $\pm$0.3 & $^{+0.8}_{-0.3}$ & $\pm$0.3 & $^{+0.4}_{-0.3}$ &  
$^{+0.30}_{-0.32}$ & $^{+0.6}_{-0.5}$ & $^{+1.5}_{-1.}$\\ \\\hline
${r_{\rm 500}}$ (EdS) & 0.723 & 0.757 & 0.867 & 1.024 & 1.041 & 0.982 & 0.879 &1.079\\ \\\hline
$f_{\rm gas}$ (EdS) & 0.0782$^{+0.009}_{-0.010}$ & 0.0987$^{+0.011}_{-0.009}$  & 0.0959$^{+0.005}_{-0.015}$  & 0.1054$^{+0.006}_{-0.005}$  & 0.0703$^{+0.005}_{-0.005}$  & 0.095$^{+0.005}_{-0.005}$  & 0.1162$^{+0.014}_{-0.013}$  & 0.1016$^{+0.025}_{-0.023}$ \\ \\ \hline
${r_{\rm 500}}$ ($\Lambda$CDM) & 1.315 & 1.358 & 1.572 & 1.87 & 1.843 & 1.799 & 1.591 & 1.911\\\\\hline
$f_{\rm gas}$ ($\Lambda$CDM) & 0.1109$^{+0.014}_{-0.015}$  & 0.1500$^{+0.018}_{-0.014}$  & 0.1431$^{+0.021}_{-0.025}$  & 0.1495$^{+0.009}_{-0.007}$ & 0.1007$^{+0.007}_{-0.008}$  & 0.1446$^{+0.009}_{-0.007}$  & 0.1825$^{+0.025}_{-0.022}$  & 0.1424$^{+0.038}_{-0.034}$  \\\\\hline  
\end{tabular} 
\caption{\label{tab:fg} Apparent gas mass fractions at ${r_{\rm 500}}$ in both EdS and $\Lambda$CDM cosmologies with $h = 0.5$, uncorrected for clumping,  with uncertainties from temperature uncertainties. Temperature measurements are taken from  Lumb et al. (2004) without cooling flow excision.} 
\end{table*} 

\subsection{The shape of the apparent gas mass fraction}

In order to investigate the global shape of $f_{\rm gas}$ profile we
have followed the procedure similar to RSB00. We compute $f_{\rm gas}$
up to the maximum radius of detection (published in Lumb et
al. 2004). Beyond this limit, this limit $f_{\rm gas}$ is obtained by
extrapolating up to the virial radius.  The radial distribution in the
local sample is derived using the published $f_{\rm gas}$ values up to
the X-ray limiting radius $R_{\rm X\,lim}$ (RSB00), upward of this radius
we computed $f_{\rm gas}$ at the two fiducial radii $R_{1000}$ and
$R_{2000}$ defined by Vikhlinin, Forman \& Jones (1999, hereafter
VFJ99) for which they published the gas masses ($Mg =
4/3\pi\rho_g(1+\Delta)R_\Delta^3(1+z)^3$), allowing a comparison in
the most outer regions.  For this comparison we do not correct for 
clumping (Mathiesen et al. 1999) as the radial variation of this
quantity is unknown. Moreover, if the scaling  holds, the
emission of both local and distant clusters should be biased by the
same amount by this effect, still allowing a meaningful direct 
comparison.

The scaled $f_{\rm gas}$ radial profiles of the individual high-z
 clusters and the mean $f_{\rm gas}$ profile of the local sample
 derived for both EdS and $\Lambda$CDM models are displayed in
 Fig. \ref{fig:fbshape}.  For a given value of the normalization
 $A_{TM}$, the virial radius (for a given temperature) depends on the
 cosmology through equation \ref{eq:rv} and \ref{eq:tm}. However, in
 practice for a NFW profile the masses inferred in a fixed physical
 radius for low redshift clusters are very similar in both
 cosmologies.  As expected, a noticeable difference in the amplitude
 of $f_{\rm gas}$ in distant clusters appears, depending on the
 cosmological model. For both cosmological models, the scaled $f_{\rm
 gas}$ profile of distant clusters is globally in good agreement with
 what has been inferred for clusters at low redshift by SB01: the
 apparent mean gas profile of our distant clusters increases from the
 center to outer shells following roughly a universal gas mass
 fraction shape. Interestingly, these $f_{\rm gas}$ exhibit a level of
 dispersion consistent with what has been found
 previously (RSB00, VFJ99).  In the EdS model the most central values
 of $f_{\rm gas}$ seem smaller in high redshift clusters. Such
 a deviation is consistent with the measured evolution of the
 {\hbox{$L_{\rm X}$--$T$} } relation, weaker than expected if the
 scaling  strictly hold. Conversely, in the low density
 flat model, $f_{\rm gas}$ values in the central parts of distant
 clusters seem to agree more with the scaling, again in agreement with
 the evolution of the {\hbox{$L_{\rm X}$--$T$} } relation in this
 cosmology. However, in the outer regions the mean $f_{\rm gas}$ in
 the $XMM-Newton$ distant sample seems not to match the
 local one very well. Examination of the average $f_{\rm gas}$ compared to the
 local one more clearly reveals a systematic difference: the inner
 mean gas mass fraction in distant clusters does not rise in as
 rapidly as in the local sample. It is unclear whether this difference
 is real, given the various origins of clusters used in the local
 sample.

\subsection{f$_{gas}$--Tx correlation}

In order to understand whether the above difference is meaningful,
an accurate knowledge of the gas mass fraction is needed. By examination of
the RSB01 sample, restricted to clusters for which the actual X-ray extension 
was known, we found a clear trend of $f_{\rm gas}$ increasing 
with temperature (this trend is much less clear in the global sample).
However, this sample was not designed to offer a uniform sample for X-ray 
studies, therefore it has been used only as a guideline in the present study. 
To examine in a more systematic and uniform way 
whether $f_{\rm gas}$  varies with temperature, 
we have computed $f_{\rm gas}$ of nearby clusters at the two fiducial radii 
$R_{2000}$, $R_{1000}$  
used in VFJ99 as well at the virial radius. 
The gas mass fraction in our high-z cluster sample has been 
estimated at the same average radii (in units of
the virial radius) allowing a direct comparison between the local and distant $f_{\rm gas}$ values.  In our distant clusters, X-ray emission is detected 
up to a radius comparable to $R_{1000}$ so direct comparison is meaningful.
Emission has to be extrapolated up to $R_{\rm v}$; it was extrapolated in both samples in similar ways. For further  comparison we have
 also computed the baryon fraction for clusters within the same redshift range
from $Chandra$ data obtained by Vikhlinin et al. (2002) whose x-ray detection 
extends typically up to  the virial radius. 

\begin{figure*}[!ht]
\includegraphics[scale=0.4,angle=0]{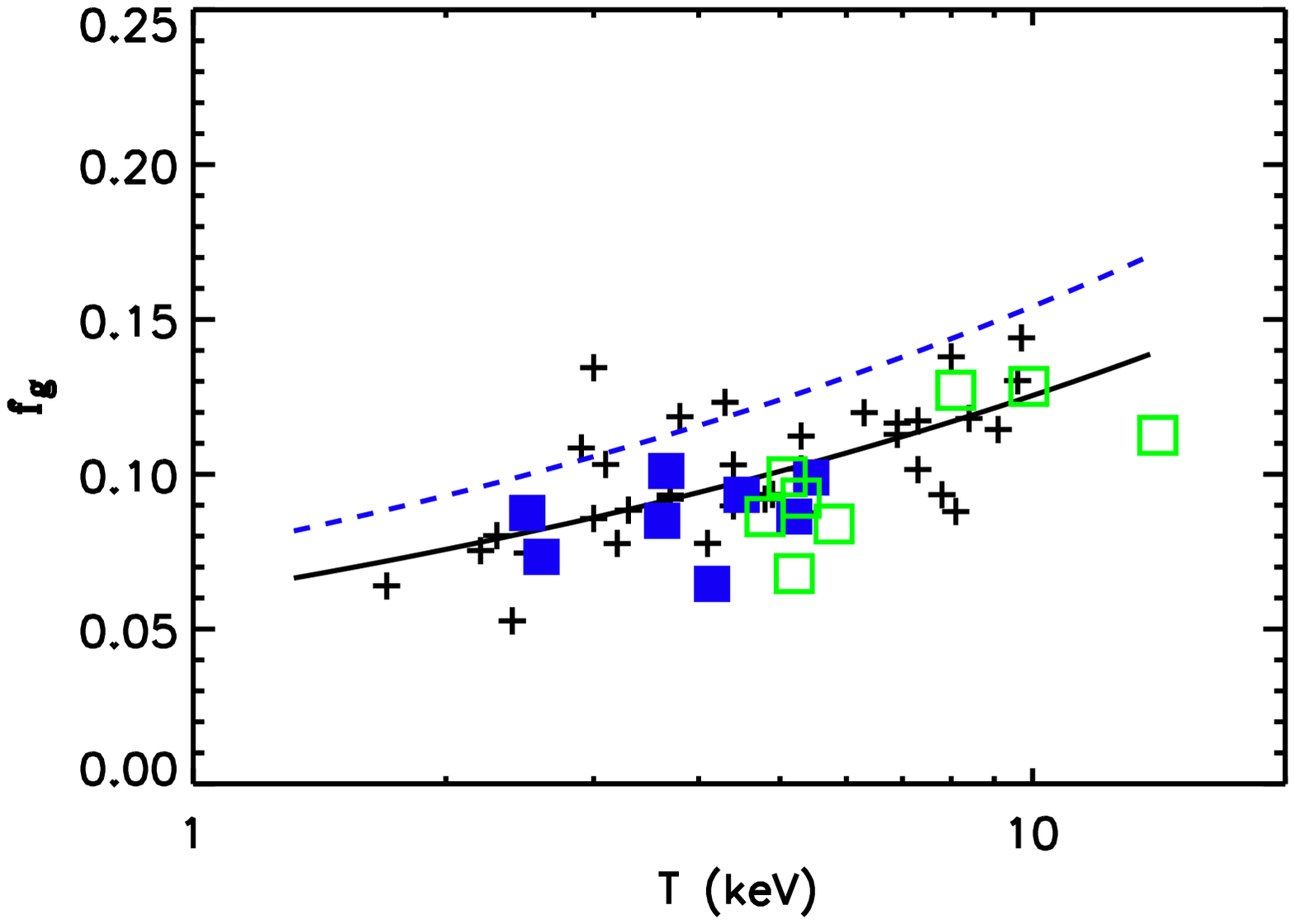}
\includegraphics[scale=0.4,angle=0]{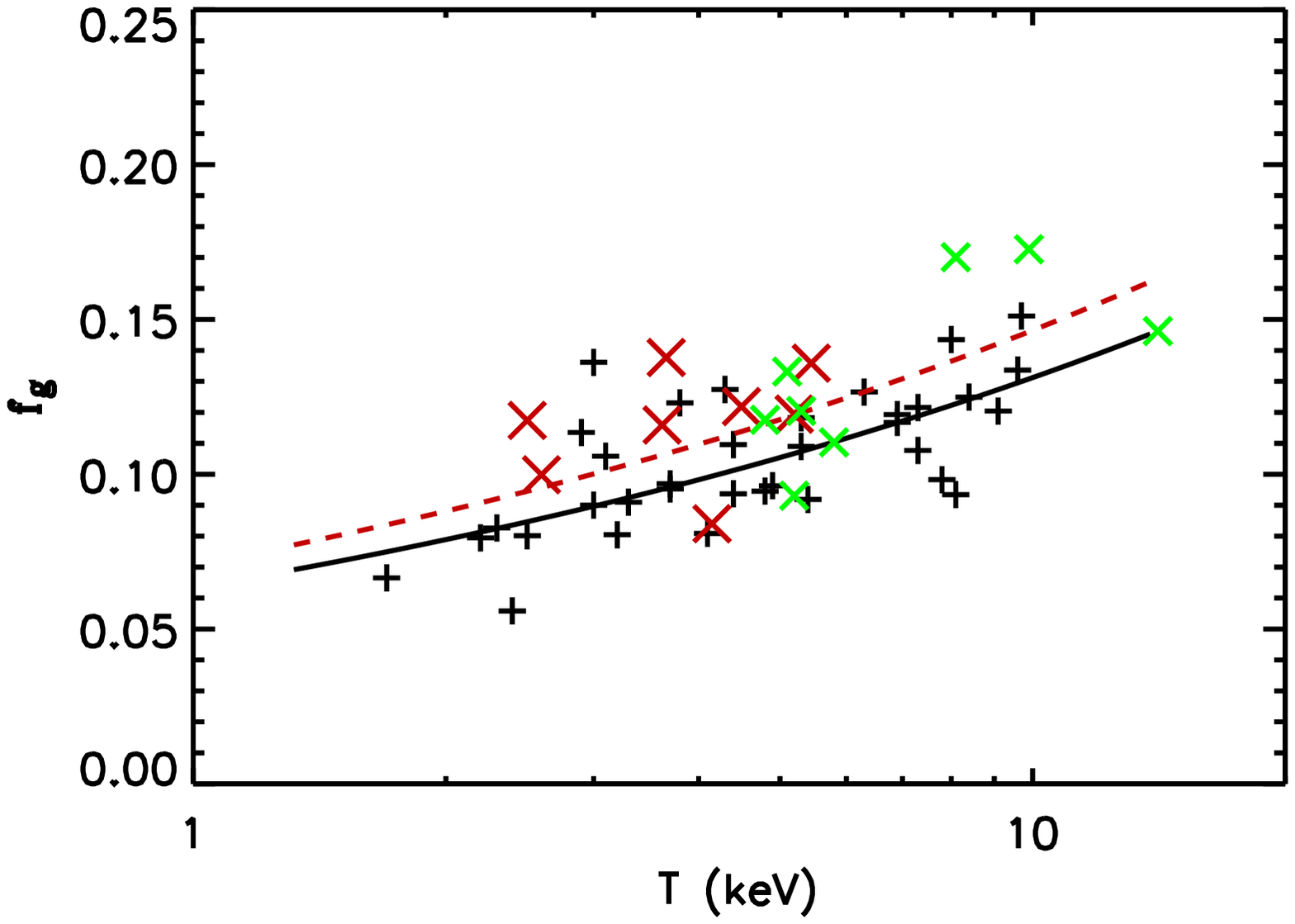}\\
\includegraphics[scale=0.4,angle=0]{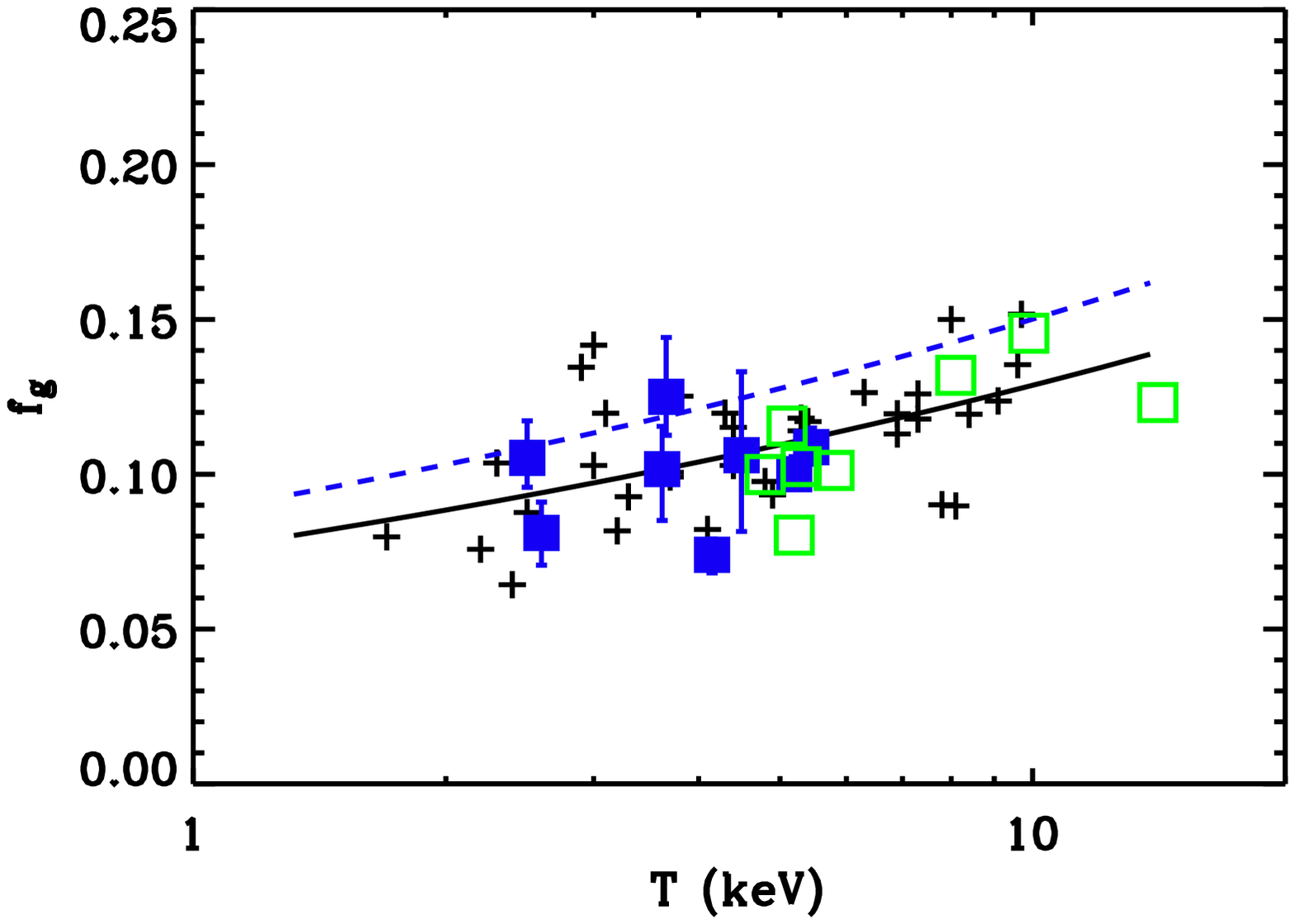}
\includegraphics[scale=0.4,angle=0]{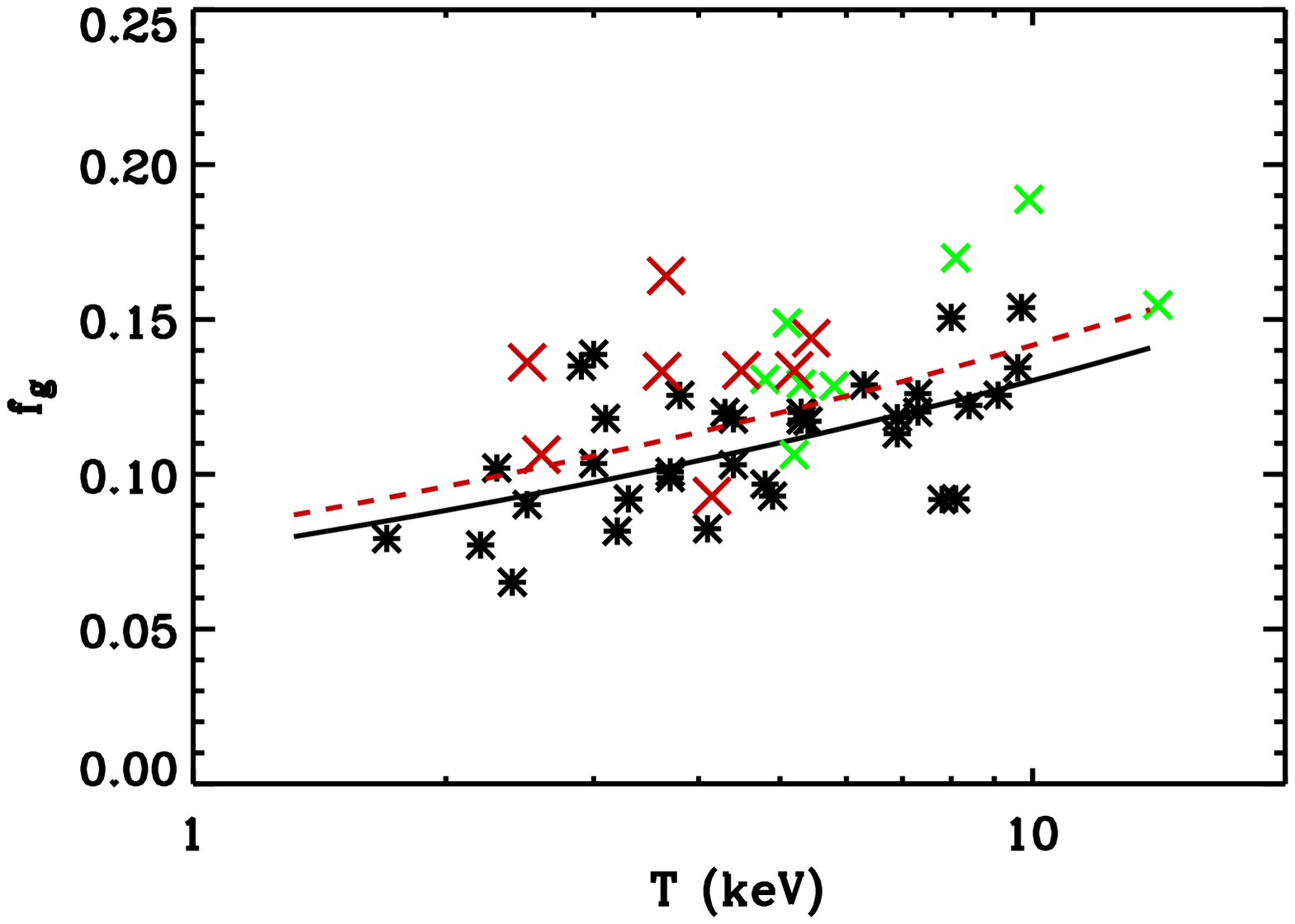}\\
\includegraphics[scale=0.4,angle=0]{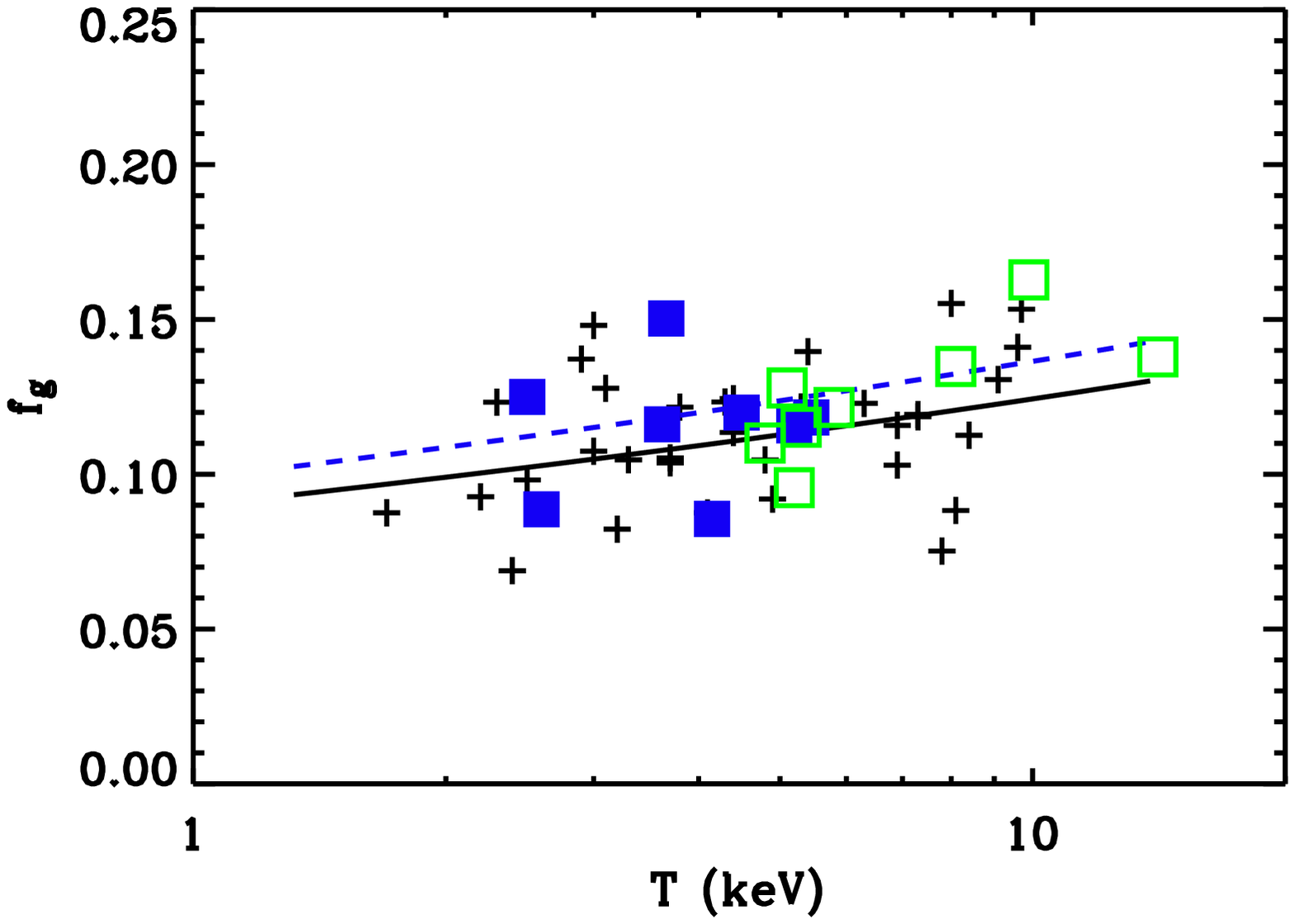}
\includegraphics[scale=0.4,angle=0]{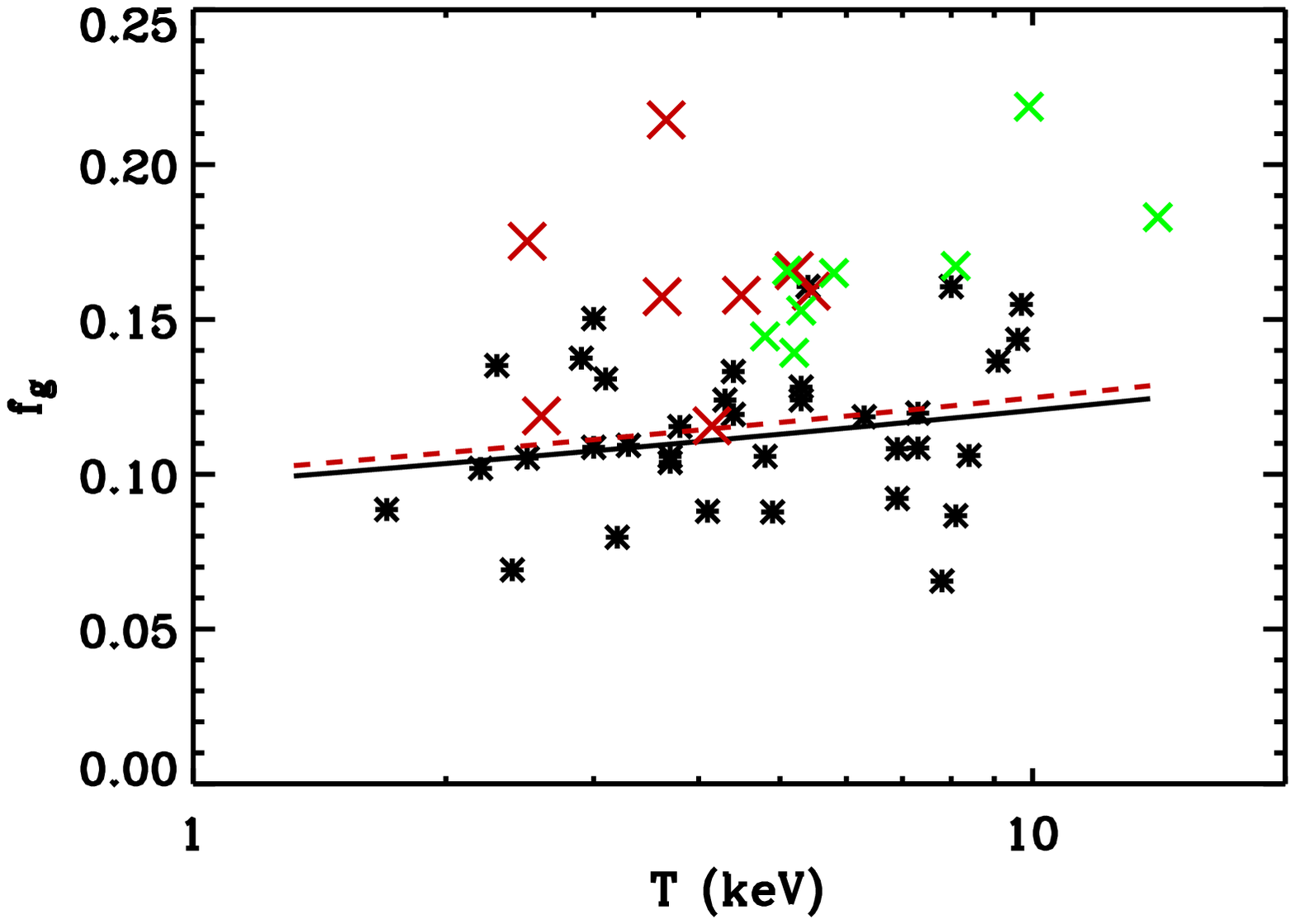}
\caption[]{ $f_{\rm gas}$ versus temperature  at three 
different radii $R_{2000}$ (top), $R_{1000}$ (middle) and $R_{\rm v}$ (bottom) in the outer parts of the $XMM-Newton$ distant clusters (blue squares and red
crosses) in Einstein de Sitter model 
(on the left) and in a concordance model (crosses on the right side) compared
to the same quantity (plus symbols) evaluated at the same scaled  radii 
{ from the   local sample  by VFJ99}. 
Open (green) squares and small (green) crosses are the same quantities
 evaluated for clusters 
in the Vikhlinin' sample {  (Vikhlinin et al., 2002)} within the same redshift range ($0.4 < z < 0.62$).
Errors bars (coming from the uncertainty on the temperature) have been drawn in one case ($R_{1000}$).
 In the concordance case,
standard scaling of the mass--temperature relation leads to gas fractions 
represented by the (red) crosses.
 Dashed (colored) lines are
the expected $f_{\rm gas}$ at the fiducial redshift of the $XMM-Newton$ clusters 
from scaling relations.}
\label{fig:fbTOM1R2}
\end{figure*}
\begin{figure*}[!ht]
\includegraphics[scale=0.4,angle=0]{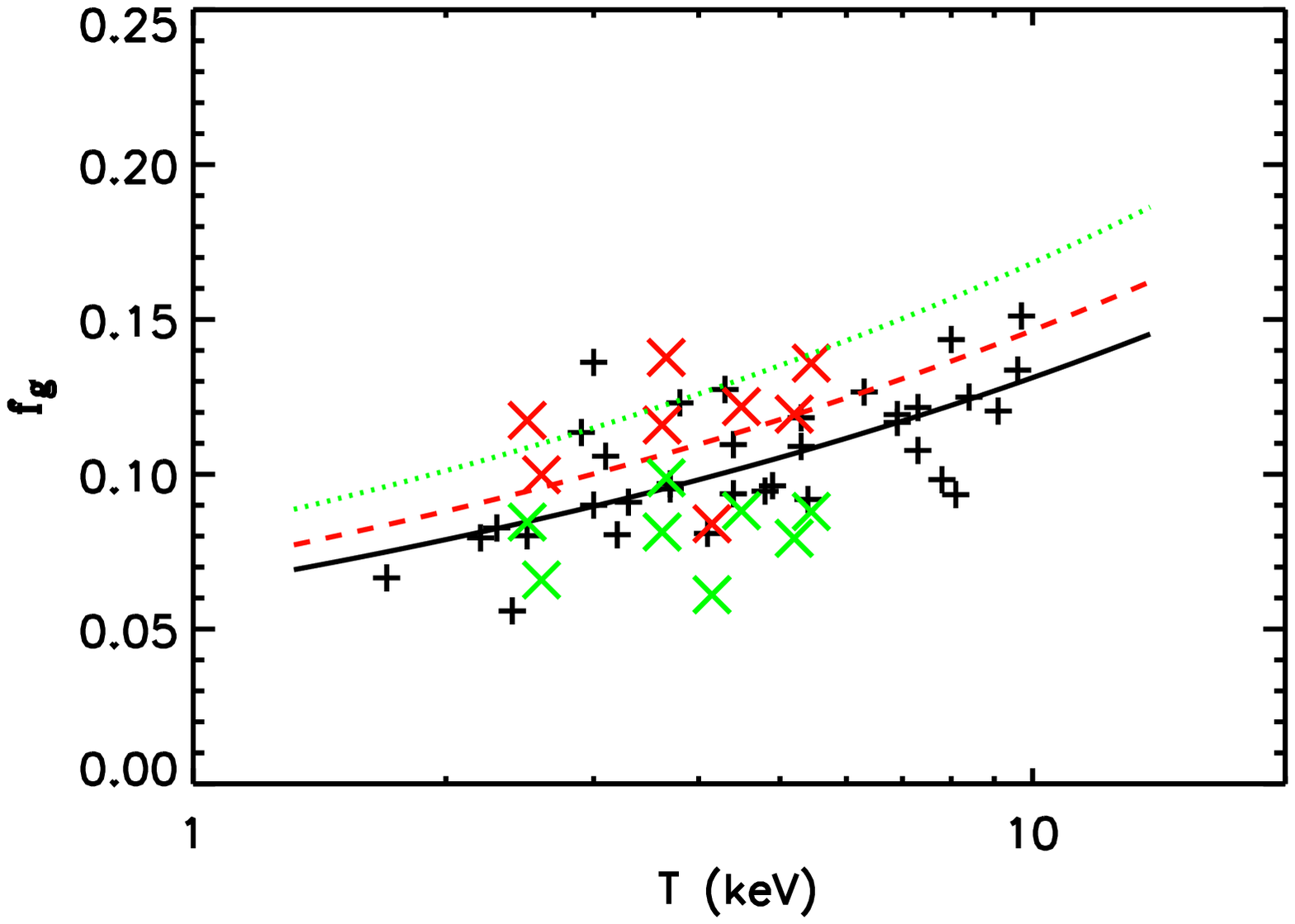}
\includegraphics[scale=0.4,angle=0]{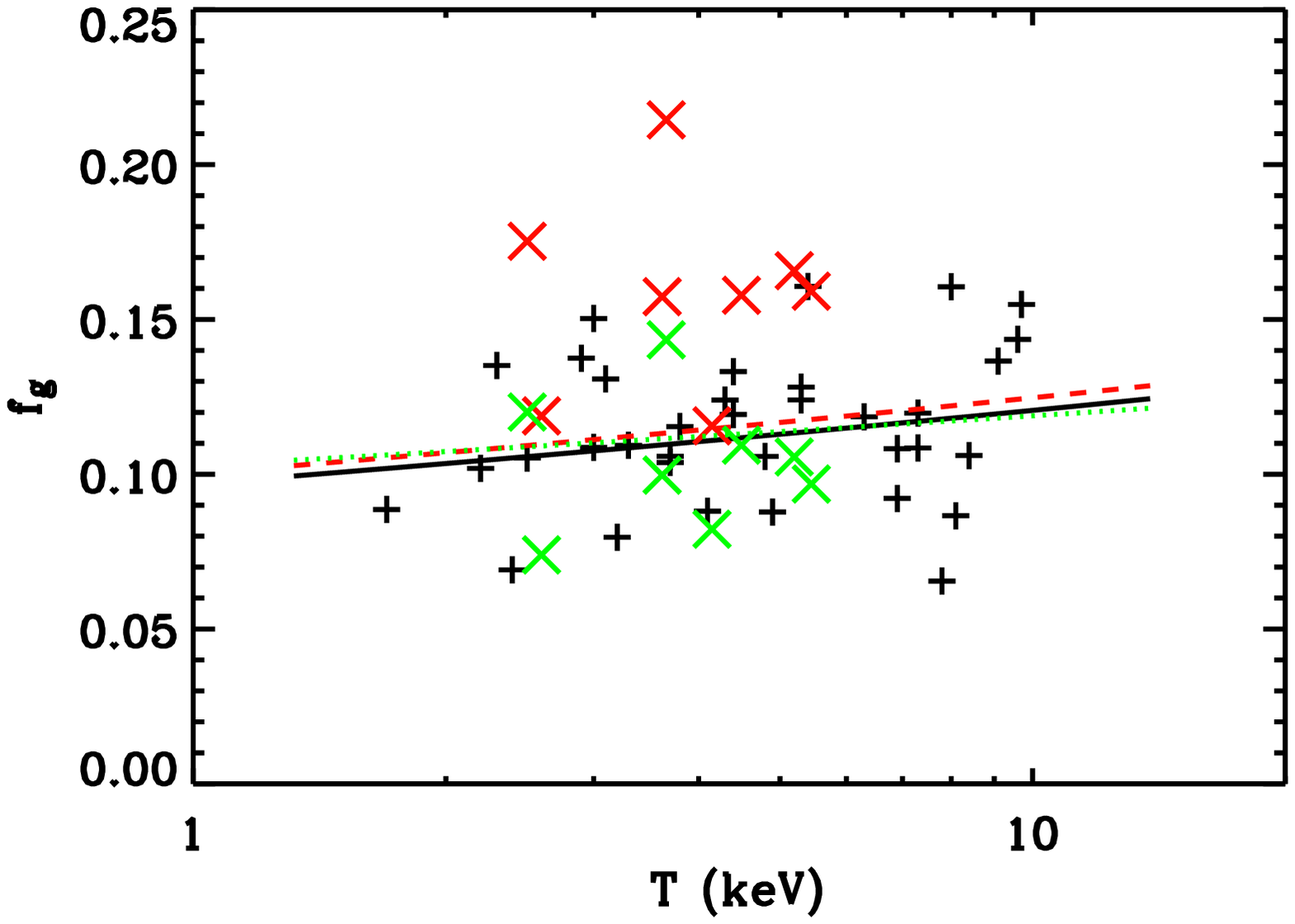}
\caption[]{ Same quantities as in Figure \ref{fig:fbTOM1R2}.
 Light grey (green) crosses were obtained for a concordance model assuming 
the non-standard scaling {\hbox{$M_{\rm V}$--$T$} } relation from Vauclair et al (2003), Eq. 4. The
dotted (green) line is the expectation from scaling in this case. Left side is at ($R_{2000}$), right side is at the virial radius.}
\label{fig:fbTME}
\end{figure*}

\begin{table}
\begin{center}
\begin{tabular}{lcc} \hline
\hline
 Radius   & Gas fraction  &  Gas fraction   \\
    &  ($\Omega_m = 1$)  & ($\Omega_m = 0.3$)   \\\hline
 $R_{2000}$      &  $0.061T^{0.31\pm0.06}$ (17\%)         &   $0.063T^{0.32\pm0.06}$ (16\%) \\ \hline
 $R_{1000}$         &   $0.075T^{0.23\pm0.06}$  (17\%)        &   $0.075T^{0.24\pm0.06}$ (17\%)\\ \hline
 Virial     &  $0.090T^{0.14\pm0.07}$  (18.8\%)      &   $0.097T^{0.095\pm0.08}$ (22\%)  \\\hline

\end{tabular}
\end{center}
\caption{Least square fit of the gas fraction to the local sample of
x-ray clusters. Uncertainty in the normalization constant is 4-5\%. The
dispersion around the fit is also given in percent. }
\label{tab2}
\end{table}

For a fixed value of the normalization $A_{TM}$ of the
mass--temperature relation, the virial radius and the total mass
enclosed in a given physical radius depend on the cosmological model.
In Figure \ref{fig:fbTOM1R2}, we have plotted $f_{\rm gas}$ in the local and high
redshift clusters versus temperature for the two cosmological models.
Results from least square fits of the local data
are given in Table \ref{tab2}, as well as 
dispersions around the best fit line. 
 $f_{\rm gas}$ values derived from the local sample
reveal a clear trend with temperature:   the gas mass fraction 
 at a fixed { scaled} radius increases
with temperature.  This trend is stronger in
the inner radius ($R_{2000}$)  than  at the outer radius ($R_{1000}$).
 At the virial radius, the inferred gas fractions are  marginally
consistent with  gas mass fractions being independent of temperature ($\sim
2.5 \sigma$ in the Einstein--de Sitter model and $\sim 1.5 \sigma$ in
the concordance model). This shows that the apparent 
$f_{\rm gas}$ in local
clusters possesses internal structure variations correlated with
temperature, therefore the scaling in local clusters is only
approximate.  The origin of these variations is unclear: it might
either be an actual variation of $f_{\rm gas}$ in clusters, but it
might also be due to a variation of the clumping of the gas with
temperature. The  $XMM$ distant sample does not reveal any clear trend with
 temperature, due to its limited temperature range.  However,
when combined with $Chandra$ clusters, a sequence appears similar to  
the one { observed in} local clusters. 
Comparison of the high-$z$ $f_{\rm gas}$ at $R_{1000}$ and $R_{2000}$
radii reveals that this internal structure also varies with redshift:
the observed mean $f_{\rm gas}$ profile of distant clusters seems to
increase toward the outer part less rapidly than in local
clusters. This introduces an additional degree of complexity when it
comes to the description of the scaling predictions. Indeed if the gas
mass fraction varies with $T$, scaling would imply:
\begin{equation}
f_g(R/R_{\rm v},T,z) =  f_g(R/R_{\rm v},T \times T_*(0)/T_*(z),z= 0)
\end{equation}
where $T_*(z)$ is the characteristic temperature associated with a 
characteristic mass scale at the epoch $z$ 
(defined by $\sigma(M_*,z) = \rm constant$). Therefore, we plot the 
predicted variations of $f_{\rm gas}$ under this scaling assumption.
Comparing predictions from this scaling scheme at the three different radii
shows that the variations with redshift of the 
internal structure do not follow the scaling either. 
This is a clear indication that clumping arising from
hierarchical building  of clusters in a purely gravitational picture 
is not the only origin of the observed complexity. Rather, it is 
likely to originate from non-gravitational heating processes, 
whose modifications
of the internal structure of clusters are  
not expected to follow standard scaling.  It is interesting that
at the virial radius, the $f_{gas}$ values we obtained for $XMM$ and $Chandra$ clusters
are consistent and suggest that the gas fraction may not vary any longer with
temperature. As our gas quantities are  extrapolated beyond emission 
detection (although not by much, especially in the case of the $Chandra$ clusters), it 
would be important to have deeper observations of the outer 
regions of  clusters to confirm this { result, both for local and distant clusters.} \\

Finally we have also considered another possibility. Vauclair et al. (2003)
have shown that the concordance model when properly normalized 
to local cluster abundances, could not reproduce  the observed
 numbers counts of distant clusters unless the mass temperature scaling 
with redshift is modified:
\begin{equation}
T =A_{TM} M^{2/3}_{15}(\Omega_0 (1+\Delta_{\rm v})/179)^{1/3}h^{2/3} \: \rm keV
\end{equation}
(i.e. the ($1+z$) term in Eq. \ref{eq:tm} has been removed).
This changes the apparent gas mass fraction in distant clusters as well as the 
predictions of the scaling model. The gas mass fractions  on both low and high
 redshift clusters were recomputed, as was   
the expected scaled variation of $f_{\rm gas}$ with temperature.
The results are shown as green symbols and lines in the colored version of 
Figure \ref{fig:fbTME}. Again, the scaling hypothesis seems not to work well under this 
scheme, perhaps not surprisingly given that the scaling has been already 
abandoned.

\section{Cosmological application}

The idea that the actual $f_{\rm gas}$ in clusters should be universal
is the starting point of an interesting cosmological test that 
has been proposed based on  the  
apparent evolution of $f_{\rm gas}$ with redshift (Sasaki 1996; Pen, 1997; 
Cooray 1998). 
The principle of this test is 
based on the fact that  the inferred gas mass fraction from X-ray data depends 
on the 
assumed cosmology through the angular distance. Comparing  the high redshift value to what is inferred from local clusters
 provides us in this way with a new test to constrain 
the cosmological parameters.

 However, from our study, it appears that there are several
 sources of complexity when applying this test. 
A first fact that should 
be taken into account
is that $f_{\rm gas}$ varies with radius inside clusters.
This variation can be accommodated if the shapes are self-similar, by working
at identical  
 scaled radius, i.e. $R/R_{\rm v}= \rm constant$ (or in a nearly equivalent way, at similar density 
contrast). Clearly, in order to prevent any 
bias one should compare $f_{\rm gas}$  
at identical radii (in units of virial radius) up to which gas emission is 
 detected. The second problem is that the  
apparent $f_{\rm gas}$ has been found to vary
with temperature (mass). There has been some debate on the strength 
of this effect  (David et al 1995, Arnaud \& Evrard 1999, Mohr et al 
1999, RSB00), but such a possibility should be kept in mind when applied to 
cosmological purposes. What we have obtained from the analysis of the VFJ99 
sample is that in the inner part of clusters both the shape and 
the amplitude of the gas mass fraction varies with 
temperature. In such a regime, it is unclear whether arguments 
based on the scaling hypothesis are valid. { However, our result are 
consistent with the 
hypothesis that the gas fraction is constant at the virial radius.}
It is therefore vital 
to have  better data on  the $f_{\rm gas}$ behavior in clusters { in their  outer part},  in order to reach conclusions 
of cosmological relevance. 

The fact that the dispersion in $f_{\rm gas}$ measurements in our
distant sample is similar to that obtained in local clusters is
very positive. Indeed, uncertainties on $f_{\rm gas}$ (from
uncertainties on temperature and flux) in our distant clusters
are significantly smaller than the intrinsic scatter (see Fig 2). From
our study it appears that although the global $f_{\rm gas}$ shape in
distant clusters is similar to the shape obtained at low redshift, the
complex internal structure, i.e. the variation of the gas mass
fraction with radius, with temperature and with redshift reveals
differences that cannot be described in a simple {\em scalable}
scheme. The observed variations in the central parts are clear
indications that $f_{\rm gas}$ evolution argument cannot be used in
this regime, given the present (lack of) understanding of the gas
physics in clusters. However, the fact that $f_{\rm gas}$ appears to
be almost constant against temperature at the virial radius is an
important piece of information indicating that the argument of a
non-evolving $f_{\rm gas}$ could be valid at this  radius. From
Fig. 2, we can see that $f_{\rm gas}$ derived from $XMM-Newton$
clusters as well as from $Chandra$ clusters are consistent with the values obtained in low redshift
clusters for the EdS model, while we observe a clear offset between
$f_{\rm gas}$ values in distant and nearby clusters computed in the
case of a $\Lambda$CDM model. The concordance model under
the assumption of standard scaling has been found to be ruled out at a
level of significance of more than 4 $\sigma$ { from the XMM data (6 
$\sigma$ when combined with Chandra measurements) } while the Einstein--de
Sitter model lies at better than $1 \ \sigma$.  \\
Several aspects however make the direct cosmological interpretation
difficult. First our sample is quite small and it would be
invaluable to have an extended version with a significantly larger
number of clusters. However, the fact that similar results are obtained from the 
$Chandra$ data is very encouraging. As a trend has been found with 
temperature, it is possible that similar trends exist with luminosity. 
Therefore
meaningful statistical comparisons require data from  X-ray selected cluster samples
 and as much as
possible from comprehensive analysis.  The second point  is
that a clear trend with temperature exists and could be a source of
confusion: temperatures in our sample are lower 
than in the local sample and therefore
expected average $f_{\rm gas}$ values are smaller (for the same
$R/R_{\rm v} <1$).  Comparing $f_{\rm gas}$ in our clusters with the
hottest local cluster, would have led to a systematic bias, at least
for the inner regions.
Indeed, it is expected that the brightest clusters (at a fixed
temperature and at a given redshift) would have a higher inner baryon
fraction than the average population.  Therefore it is natural that
the trend observed with temperature is also present with luminosity.
Finally, we showed that the internal structure of the gas is
not strictly identical in high and low redshift clusters, declining
faster in the central part of high redshift clusters. This is
consistent with what seems already to emerge from Fig. 1. This complex
internal structure is probably the result of non-gravitational (pre)
heating of the gas which is currently advocated to explain the
observed {\hbox{$L_{\rm X}$--$T$} } relation,  but might also result from
more fundamental departure of scaling laws in the dark matter, for instance 
if $c$ evolved with redshift.  These systematic
variations with radius, temperature and redshift imply that the
baryon fraction test should be performed with caution and probably
only at the virial radius, although  one expects that
outer regions are more affected by clumping and it is not clear
that clumping should follow simple scaling relations in a low density
universe.\\

 \section{Conclusion}

The observations of distant clusters with $XMM-Newton$ offer an unique
possibility to investigate the outer emissivity of the gas
distribution in distant X-ray clusters. These observations have
revealed for the first time the existence of a complex internal
structure which does not follow simple scaling laws but still does
show some regularities.  The $XMM-Newton$ observations of the distant
SHARC clusters  reveal some interesting results on this issue:
the shape of the apparent $f_{\rm gas}$ derived for these clusters is
in good agreement with the shape inferred by SB01 for local clusters.
This is an independent confirmation that the scaled shape of the gas
mass fraction in clusters is in rough agreement with numerical
simulations.  However, our analysis reveals some deviations from the standard
scaling  in the $f_{\rm gas}$ profile  of high redshift clusters.  
Furthermore, by comparing our distant clusters to a sequence
of local clusters, we found a clear variation of the internal structure 
with temperature
and redshift  which cannot
be described by simple scaling relations.  This implies that
 the
baryon fraction evolution, or lack thereof, cannot be used as a reliable
cosmological test without better understanding of the internal structure of clusters.  
Nevertheless, at the virial radius the gas fraction 
seems to be independent of temperature in the low redshift sample and
therefore that the high-redshift clusters might be used to apply the
cosmological test based on the assumption that $f_{\rm gas}$ at the
virial radius has a universal value independent of redshift,
although existing data do not allow a firm statement about this
hypothesis. From our sample of high redshift clusters at the virial
radius the data are found to be roughly consistent with a
non-evolving $f_{\rm gas}$ in an Einstein-de Sitter model, but not
within the standard concordance model in which the inferred apparent
$f_{\rm gas}$ appears systematically higher than in local clusters, unless
a non-standard scaling with redshift of the $M-T$ relation (Vauclair et al. 
2004) is used. 
However, the complex internal structure of the gas
revealed by the present analysis of $XMM-Newton$ clusters prevents 
us  from drawing definitive conclusions on cosmological parameters, as the relevant
quantities were extrapolated at the virial radius.  In the 
concordance model the mean $f_{\rm gas}$ value
estimated inside the radius of detections, is still found to lie at 4
$\sigma$ above the mean gas mass fraction of nearby clusters with
similar temperature.

\begin{acknowledgements}
      
 This research  made use of the X-Ray Clusters Database (BAX)
which is operated by the Laboratoire d'Astrophysique de Toulouse-Tarbes (LATT),
under contract with the Centre National d'Etudes Spatiales (CNES). We
acknowledge useful comments from the referee which contributed  improving
the content of this paper.

\end{acknowledgements}

\end{document}